\documentclass[fleqn,usenatbib]{mnras}

\usepackage{newtxtext,newtxmath}
\usepackage[T1]{fontenc}
\DeclareRobustCommand{\VAN}[3]{#2}
\let\VANthebibliography\thebibliography
\def\thebibliography{\DeclareRobustCommand{\VAN}[3]{##3}\VANthebibliography}

\usepackage[normalem]{ulem}
\usepackage{graphicx, subfigure, algorithm, xcolor, algorithmic, amsmath, amssymb, float, rotating, enumerate}

\newcommand{\rb}{\mbox{$\rm \text{R}_\text{birth}$}}
\newcommand{\rguide}{\mbox{$\rm \text{R}_\text{guide}$}}
\newcommand{\mgfe}{\mbox{$\rm [Mg/Fe]$}}

\newcommand{\feh}{\mbox{$\rm [Fe/H]$}}
\newcommand{\xfe}{\mbox{$\rm [X/Fe]$}}

\newcommand{\smg}{\mbox{$\rm [{\it s}/Mg]$}}

\newcommand{\fehC}{\mbox{$\rm [Fe/H](R = 0, \tau$)}}
\newcommand{\gradFeh}{\mbox{$\rm \nabla [Fe/H](\tau)$}}

\newcommand{\sh}{\textsl{\textsc{StarHorse}}}
\newcommand{\logg}{\mbox{$\log g$}}
\newcommand{\teff}{\mbox{$T_{\rm eff}$}}
\newcommand{\rgal}{\mbox{$\rm R_\text{gal}$}}

\newcommand{\numApp}{36,652}
\newcommand{\numGal}{24,467}

\graphicspath{{./}{plots/}}

\title[The \smg--age--\rb\ relation]{Chemical clocks and their time zones: understanding the [s/Mg]--age relation with birth radii}

\author[Ratcliffe, Minchev, Cescutti, et al.]{Bridget Ratcliffe,$^{1}$\thanks{E-mail: bratcliffe@aip.de}
Ivan Minchev,$^{1}$
Gabriele Cescutti,$^{2,3,4}$ 
Emanuele Spitoni,$^{3}$
Henrik J\"{o}nsson,$^{5}$ \newauthor
Friedrich Anders,$^{6, 7,8}$
Anna Queiroz,$^{1,9,10}$
Matthias Steinmetz$^{1}$
\\
$^{1}$Leibniz-Institut f\"{u}r Astrophysik Potsdam (AIP), An der Sternwarte 16, 14482 Potsdam, Germany\\
$^{2}$Dipartimento di Fisica, Sezione di Astronomia, Università di Trieste, via G. B. Tiepolo 11, 34143 Trieste, Italy\\
$^{3}$INAF. Osservatorio Astronomico di Trieste, via G.B. Tiepolo 11, 34131, Trieste, Italy\\
$^{4}$ INFN, Sezione di Trieste, Via A. Valerio 2, I-34127 Trieste, Italy \\
$^{5}$ Materials Science and Applied Mathematics, Malm\"{o} University, SE-205 06 Malm\"{o}, Sweden\\
$^{6}$Dept. de Física Quàntica i Astrofísica (FQA), Universitat de Barcelona (UB), C Martí i Franqués, 1, 08028 Barcelona, Spain\\
$^{7}$Institut de Ciències del Cosmos (ICCUB), Universitat de Barcelona (UB), C Martí i Franqués, 1, 08028 Barcelona, Spain\\
$^{8}$Institut d’Estudis Espacials de Catalunya (IEEC), C Gran Capità, 2-4, 08034 Barcelona, Spain\\
$^{9}$Laboratório Interinstitucional de e-Astronomia - LIneA, Rua Gal. José Cristino 77, Rio de Janeiro, RJ - 20921-400, Brazil\\
$^{10}$Institut f\"{u}r Physik und Astronomie, Universität Potsdam, Haus 28 Karl-Liebknecht-Str. 24/25, D-14476 Golm, Germany
}

\date{Accepted XXX. Received YYY; in original form ZZZ}

\pubyear{2023}

\begin{document}
\label{firstpage}
\pagerange{\pageref{firstpage}--\pageref{lastpage}}
\maketitle

\begin{abstract}
The relative enrichment of s-process to $\alpha$-elements ([s/$\alpha$]) has been linked with age, providing a potentially useful avenue in exploring the Milky Way's chemical evolution. However, the age--[s/$\alpha$] relationship is non-universal, with dependencies on metallicity and current location in the Galaxy. In this work, we examine these chemical clock tracers across birth radii (\rb), recovering the inherent trends between the variables. We derive \rb\ and explore the [s/$\alpha$]--age--\rb\ relationship for \numApp\ APOGEE DR17 red giant and \numGal\ GALAH DR3 main sequence turnoff and subgiant branch disk stars using [Ce/Mg], [Ba/Mg], and [Y/Mg]. We discover that the age--\smg\ relation is strongly dependant on birth location in the Milky Way, with stars born in the inner disk having the weakest correlation. This is congruent with the Galaxy's initially weak, negative \smg\ radial gradient, which becomes positive and steep with time. We show that the non-universal relations of chemical clocks is caused by their fundamental trends with \rb\ over time, and suggest that the tight age--\smg\ relation obtained with solar-like stars is due to similar \rb\ for a given age. Our results are put into context with a Galactic chemical evolution model, where we demonstrate the need for data-driven nucleosynthetic yields.
\end{abstract}

\begin{keywords}
Galaxy: abundances -- Galaxy: evolution -- Galaxy: disc -- stars: abundances 
\end{keywords}

\section{Introduction} \label{sec:intro}

% \cite{DOrazi2009, Maiorca2011} found that young stars are higher in s-process than old ones

% \cite{Jofre2020} - c/ba shows strong trend with age

Chemical abundances are connected to a star's age and place of birth \citep{2002freeman-BH, 2022Ratcliffe}, providing an effective tool in analyzing the Milky Way's evolutionary history. In particular, recent works have utilized the link between abundances and age to estimate stellar ages for large samples of red giant branch stars \citep{Hayden2022, Ciuca2023, Leung2023, Anders2023_ages}, where traditional methods fail or only provide ages for small samples in select fields. While these abundance-driven ages have improved our understanding of important evolutionary events, some of the models can be quite complex, and may not capture the precise nature of the abundance--age relationship.

Over the past few years, the ratio between s-process and $\alpha$-elements ([s/$\alpha$]) has been found to be linearly related with age in Milky Way solar-like stars \citep[][]{Nissen2015, Nissen2020} and giants \citep{Slumstrup2017}, as well as in other galaxies \citep{Skuladottir2019}. Understanding this link between [s/$\alpha$] and age could provide a way to examine chemical evolution by omitting the dependencies in modeling age. The negative correlation between [s/$\alpha$] and age is driven by the differing timescales to synthesize the two elements, as $\alpha$-elements are created during supernovae II (SNII) which happen on fast timescales, while s-process elements are produced in the asymptotic giant branch (AGB) phase of low- and intermediate-mass stars which happens on a slower timescale (\citealt{Karakas2014}, \citealt{Matteucci2021} and references within). The relative differences in the timescales suggests the possibility that the [s/$\alpha$]\footnote{The most common chemical clock is [Y/Mg] or [Y/Al], though similar trends are found across other s-process and $\alpha$-elements.} abundance can be an age indicator, or chemical clock \citep{Nissen2015, TucciMaia2016, Spina2016}. 

Using only 25 solar-like stars, \cite{daSilva2012} found a correlation between [s/$\alpha$] and age, which has successfully been replicated with more stars in other works using solar-type stars in the solar neighborhood \citep{Nissen2015, Nissen2016, TucciMaia2016, Titarenko2019, Nissen2020, Jofre2020}. However, its dependency on location in the Galaxy (\citealt{Casali2020, Casamiquela2021, ViscasillasVazquez2022}) and metallicity (\citealt{Feltzing2017, DelgadoMena2017, SalesSilva2022}) is an ongoing question. Quantifying this radial dependency is a non-trivial task, as stars radially migrate away from their birth sites \citep{Selwood2002, Roskar2008, 2009schonrichBinney, Minchev2010, Frankel2020}, blurring fundamental chemical relations \citep[e.g.][]{Ratcliffe2023_enrichment}. 

\begin{figure*}
     \includegraphics[width=.85 \textwidth]{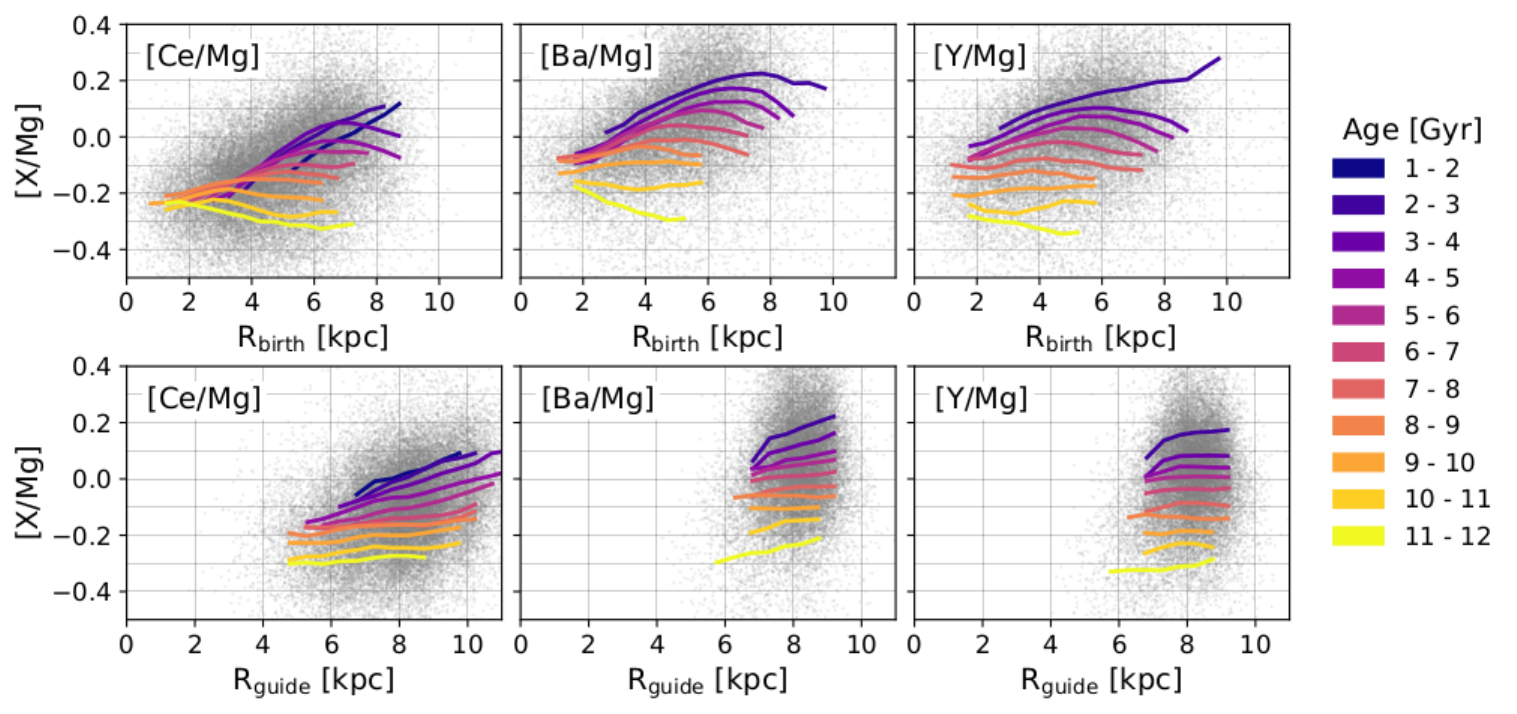}
\caption{Time evolution of \smg\ abundance gradients for our APOGEE ([Ce/Mg]) and GALAH ([Ba/Mg], [Y/Mg]) samples. \textbf{Top:} Running means of different mono-age populations, determined by calculating the average \smg\ for \rb\ bins of width 0.5 kpc and then smoothed over 1 kpc. Note that age here becomes look-back time, since we are using \rb. \textbf{Bottom:} Same as the top panels, but looking at the variation over age with \rguide. \smg\ has a weak relationship with \rb\ for older aged stars, which becomes positive and stronger with decreasing lookback time. The outer disk sees a faster enrichment in \smg\ than the inner disk due to AGB stars migrating outwards before polluting the interstellar medium.} 
\label{fig:xmg_R}
\end{figure*}

Recently, works have utilized the homogeneity of birth clusters \citep{BH2010} to estimate stellar birth radii (\rb) directly from their metallicities and ages. In order to assign a specific \feh\ and age to a birth location, the relationship between age, birth radius, and \feh\ over time needs to be known. While this relationship can be modeled \citep{Frankel2018}, more recent works have proposed methods which require less assumptions. In order to derive \rb, \cite{Minchev2018_rbirth} simultaneously recovered the metallicity evolution in the Galactic disk over time by enforcing the resulting \rb\ distributions to remain physically meaningful. More recently, \cite{Lu2022_Rb} discovered a linear relation between the scatter in \feh\ for a given age and the \feh\ gradient evolution with time, allowing for the evolution of the metallicity gradient to be recovered with minimal assumptions. Applying this method to APOGEE DR17, \cite{Ratcliffe2023_enrichment} derived \rb\ for 145,447 red giant stars to recover the time evolution of chemical abundances across the Milky Way disk, revealing the information lost when only considering the radial gradient evolution estimated using current radius. That work put in context (and quantified) the expectation from models and simulations that radial mixing has significantly distorted present day chemo-kinematical relations used for inferring the Milky Way formation history \citep[e.g.][]{Pilkington2012, Minchev2012, Minchev2014, Kubryk2013, Vincenzo2020}.

In this work, we explore the universality of chemical clocks across birth radii for giants and main-sequence turnoff and subgiant branch (MSTO+SGB) disk stars. In particular, we examine how the age---\smg\ relationship differs across the Galaxy after minimizing the effect of radial migration. Section \ref{sec:data} presents the data sets used in this analysis. Section \ref{sec:results} gives the results and discussion of our findings in relation to previous works. Our conclusions are given in Section \ref{sec:conclusion}.

\section{Data} \label{sec:data}

We examine three s-process elements --- Ce, Ba, and Y --- to study the relationship between [s/Mg], age, and birth radii. We use the [Ce/Fe] abundance from the BACCHUS Analysis of Weak Lines in APOGEE Spectra (BAWLAS; \citealt{Hayes2022_bawlas}) catalog, and pair it with \feh\ and \mgfe\ abundances from APOGEE DR17 \citep{apogeeDR17, Majewski2017} of the fourth phase of the Sloan Digital Sky Survey (SDSS-IV; \citealt{blanton2017sloan}), which are processed using the APOGEE Stellar Parameter and Chemical Abundance Pipeline (ASPCAP; \citealt{Holtzman2015, GarciaPerez2016aspcap, Jonsson2020}). We partner these abundances with spectroscopic stellar ages from \cite{Anders2023_ages} and guiding radii (\rguide) calculated as: $$\rguide = \rgal V_\phi / V_0,$$
where $V_\phi$ and \rgal\ are the Galactocentric azimuthal velocity and Galactocentric cylindrical radius from the astroNN catalog \citep{LeungBovy2019a}, and $V_0=229.76$ km/s is the Milky Way rotation curve at solar radius \citep{Bovy2012_velocityCurve, Schonrich2010_lsr}. To ensure that we select a high quality sample, we use APOGEE red giant ($2 < \logg < 3.6$, 4\,250
$\leq \teff \leq 5\,500$ K) disk ($|z| < 1$ kpc, eccentricity $<0.5$, $|\feh|< 1$) stars with unflagged abundances, $\feh_\text{err}<0.015$, age$_\text{err}< 1.5$ Gyr, $SNR > 100$, age $<12$ Gyr, |[Ce/Fe]$|\leq 1$, and [Ce/Fe]$_\text{err}<0.07$ dex. 

We additionally use the [Y/Fe], [Ba/Fe], \mgfe, and \feh\ abundance measurements provided in GALAH DR3 \citep{Buder2021}, with ages from the \sh\ value-added catalog \citep{Queiroz2023_SH}, which was recently corrected in a new version\footnote{\href{https://data.aip.de/projects/aqueiroz2023.html}{https://data.aip.de/projects/aqueiroz2023.html}}. \rguide\ is again calculated from cylindrical rotational velocity, with kinematics from the dynamics value-added catalog. For our GALAH MSTO+SGB disk ($|\feh| < 1$, eccentricity $<0.5$, $|z|<1$ kpc) sample, we perform similar cuts as above; we keep stars with unflagged abundances, $|\xfe| < 1$, $\feh_\text{err}<0.1$, age$_{84}-$  age$_{16}< 2.5$ Gyr, {\tt snr\_c2\_iraf} $>$ 50, $2 < \text{age} < 12$ Gyr, and [X/Fe]$_\text{err}<0.1$ dex.

We estimate a star's birth radius directly from their \feh\ and age measurements: 
\begin{equation} \label{eqn:rb}
\rb(\text{age}, \feh) = \frac{\feh - \fehC}{\gradFeh},
\end{equation}
using the birth metallicity gradient (\gradFeh) and central metallicity evolution over cosmic time $\tau$ (\fehC) from \cite{Ratcliffe2023_enrichment} (see their section 3 for more information on the birth radii derivation). We correct for the 0.05 dex systematic offset in \feh\ between GALAH and APOGEE \citep{Buder2021} before calculating \rb\ for our GALAH sample. Since the main goal of this work is to look at overall trends with \rb\ between \smg\ and age, and not to directly compare the numerical relationships between catalogs, we do not correct for other potential offsets. To get the \smg\ abundance, we simply subtract \mgfe\ from [$s$/Fe]. After performing the quality cuts listed above and removing stars with $\rb < 0$, we are left with \numApp\ APOGEE giants and \numGal\ GALAH MSTO+SGB stars. 

\begin{figure*}
     \includegraphics[width=.85\textwidth]{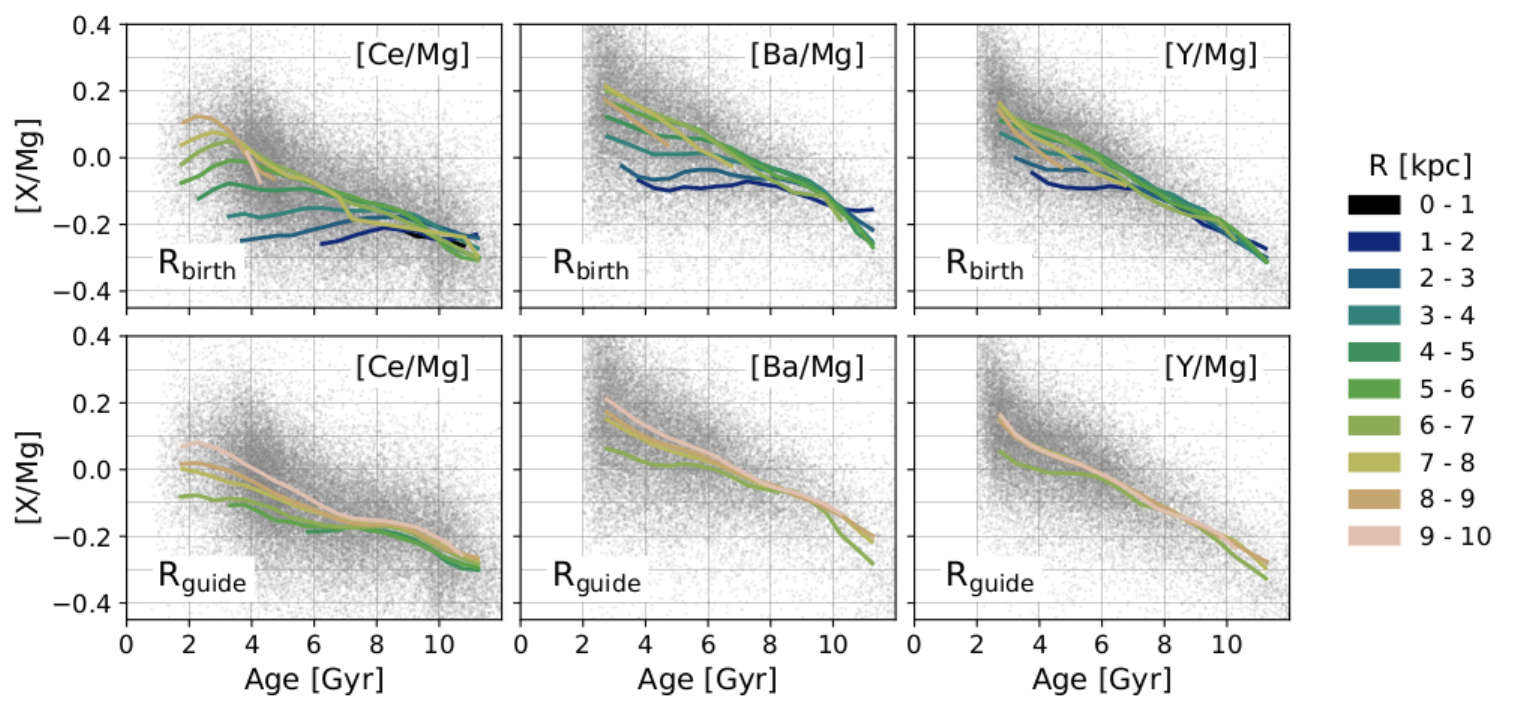}
\caption{The age--\smg\ plane of [Ce/Mg] (APOGEE), [Ba/Mg] (GALAH), and [Y/Mg] (GALAH) overlaid by the running means of {\bf top:} mono-\rb\ populations and {\bf bottom:} mono-\rguide\ populations. The running means are measured by calculating the average \smg\ for age bins of 0.5 Gyr and smoothing over 1 Gyr. The smaller \rb\ tracks show a weak relationship with age while the correlation between \smg\ and age becomes quite strong for larger \rb. Each \rb\ sees a similar trend with \smg\ over time; a radius sees an initial increase in \smg\ until it flattens and stays nearly constant with time until the present day. This trend is drastically different than looking at mono-\rguide\ groups, which show minimal differences beyond slightly steeper slopes for outer radii.} 
\label{fig:xmg_age}
\end{figure*}

\section{Results and discussion} \label{sec:results}

With estimates of stellar ages and birth radii for large samples of stars, we can analyze the properties of chemical clocks over lookback time --- which is not the same as age, see e.g. \cite{Ratcliffe2023_enrichment} --- and assess the effect of radial migration by comparing these properties to observed trends with guiding radii. 

\subsection{Radial gradient evolution}\label{sec:rad_grad}

Using our \rb\ estimates, we find distinct trends between \smg\ and \rb\ in the top panels of Figure \ref{fig:xmg_R}, where the \rb--\smg\ relation illustrates the time evolution of \smg\ abundance gradients. In particular, we find that for all three of our \smg\ relations, the youngest mono-age populations have a steep positive radial gradient, while the oldest mono-aged populations have a weaker radial gradient. In fact, the three oldest age groups correspond to the high-$\alpha$ sequence (see \citealt{Anders2023_ages, Queiroz2023_SH}), which appear to have a slightly negative \smg\ radial birth gradient. 

Once the low-$\alpha$ sequence began forming, the \smg\ radial gradient began quickly steepening with time. This steepening in the radial gradient visualized in the top panels of Figure \ref{fig:xmg_R} can be interpreted as a rapid increase in the enrichment with Galactic radius. This can be explained as an effect of radial migration, since the production of the s-process elements occurs during the AGB phase, which gives sufficient time for migration to proceed. The larger fraction of s-process elements at larger radius is then due to low- and intermediate-mass AGB stars migrating before polluting the interstellar medium. Indeed, it is well-established \citep{Minchev2014, Frankel2020} that outside $\sim1$ disk scale-length ($\sim3.5$ kpc; \citealt{BlandHawthorn2016}), stars migrate larger distances outward and contribute more to a given radius. The steep \smg\ radial gradient for the youngest populations also explains the scatter in \smg\ found in young ($<2$ Gyr) open clusters \citep{PenaSuarez2018, Casali2020}, as there is a range of \smg\ values throughout the disk for a given (younger) age.

The bottom row of Figure \ref{fig:xmg_R} presents the \rguide--\smg\ plane for mono-age populations, illustrating the differences in measuring the radial \smg\ gradients with \rb\ (top row) and \rguide\ (bottom row). Similar to \cite{Ratcliffe2023_enrichment}, we find that the observed radial gradients with \rguide\ are weaker than the radial birth gradients, showing the level to which radial migration masks fundamental trends. Particularly for [Ba/Mg] and [Y/Mg] (GALAH sample), the \rguide--\smg\ relation has a minor correlation overall. [Ce/Mg] (APOGEE sample) shows similar trends, however due to the larger spatial coverage, the younger aged populations show a positive radial slope which is only hinted at in the GALAH sample.

\subsection{Enrichment with cosmic time}

While the \smg\ radial gradient can provide insights into how the relative fraction of s-process to $\alpha$-elements varies across the disk with time, the age--\smg\ relation is potentially a useful tool in estimating stellar age that needs better understanding. Previous works have found that the age--[s/$\alpha$] relation changes as a function of current radius, suggesting that the differences arise from the strong, non-monotonic dependence of the s-process yields with metallicity and different star formation histories \citep{Casali2020}. However, due to radial mixing processes, a given location in the Galaxy is comprised of stars born in a range of Galactic radii (depending on the age; see, e.g. \citealt{Minchev2018_rbirth, Agertz2021_vintergatanI}).

The top panels of Figure \ref{fig:xmg_age} show the age--\smg\ relation for stars born at different locations in the Galaxy. We find that most mono-\rb\ populations increase in \smg\ before flattening out and seeing minimal differences in \smg\ at later times. The flattening occurs at increasingly younger age for larger radii, which can be linked to the disk inside-out growth. In our data sets, the larger \rb\ bins do not show a weakening in the age--\smg\ gradient due to the lack of younger stars. This trend is consistent across red giants ([Ce/Mg]) and MSTO+SGB stars ([Ba/Mg], [Y/Mg]), and is in disagreement with previous work that found the age--\smg\ relation to be weaker in field giants and stronger in dwarfs \citep{KatimeSantrich2022}. This shows that once working with \rb, the differences among different stellar evolutionary states is minimized. Given that each birth radius has a similar trend in enrichment of \smg, this suggests that the effect of star formation is simply to shift the trends to larger radii as the disk grows with time.

In contrast to the top panels of Figure \ref{fig:xmg_age}, the bottom row shows the age--\smg\ relation for stars in different guiding, rather than birth, radial bins. We find that the well-defined self-similar trends seen for the mono-\rb\ populations are now strongly blurred and largely overlapping. In particular, the flattening for younger ages at a given radius is almost completely lost, especially for [Ba/Mg] and [Y/Mg], except for a hint in the innermost \rguide\ bin.

Using open clusters, \cite{ViscasillasVazquez2022} found that [Ba/Mg] and [Y/Mg] have respectively a weak and inverted trend with age in the inner disk. Using birth radii, we find that both [Ba/Mg] and [Y/Mg] have a weakly positive relation with age for smaller \rb. In fact, we find the mono-\rb\ populations behave similarly for all 3 elements (albeit with different slopes) in the \smg--age plane, which differs from previous findings reporting that [Y/Mg] and [Ba/Mg] have different trends (e.g. \citealt{daSilva2012}), and shows the additional information gained using birth radii. Recently, \cite{Casali2023} showed that stars located in outer regions of the disk lie on steeper slopes in the age--[Ce/$\alpha$] plane than stars currently located more inwards. Our work agrees with this finding, with our Figure \ref{fig:xmg_R} illustrating that this is caused by a more intense enrichment of \smg\ in the outer disk. 

As mentioned in Section \ref{sec:rad_grad}, the \smg\ radial gradient has been strengthening with time. In both our GALAH and APOGEE samples, the valley between the high- and low-$\alpha$ sequences happens between 9 and 10 Gyr ago, suggesting that chemical clocks behave differently for the two sequences. The validity of chemical clock indicators using s-process elements in the high-$\alpha$ sequence has been questioned before, with arguments that AGB stars would not have produced enough s-process elements that quickly \citep{Hayden2022}. However, we see in Figure \ref{fig:xmg_age} that these older stars have the strongest correlation with age, showing minimal differences across the different mono-\rb\ populations. This also contrasts the differences between using current and birth radius, as the age--\smg\ relation has been found to be flat for older stars \citep{Casali2023}. One way to interpret this is that as the disk evolved, factors --- such as radial migration processes --- caused a variation in the production of \smg\ as a function of radius, therefore creating a spread in \smg\ abundance for a given age which increased with time. We explore these factors in Section \ref{sec:model} using Galactic chemical evolution modeling.

% One way to interpret this is that the initial burst of SNII polluting the ISM with $\alpha$-elements created stars with a similar amount of \smg\ abundance throughout the early disk. As the disk evolved, other factors, such as a non-uniform initial mass function and radial migration processes caused a variation in the production of \smg\ as a function of radius, therefore creating a spread in \smg\ abundance for a given age.

\subsection{Tight age--[s/Mg] relationship in solar twins is driven by varying birth radii}

\begin{figure}
    \centering
     \includegraphics[width=.37\textwidth]{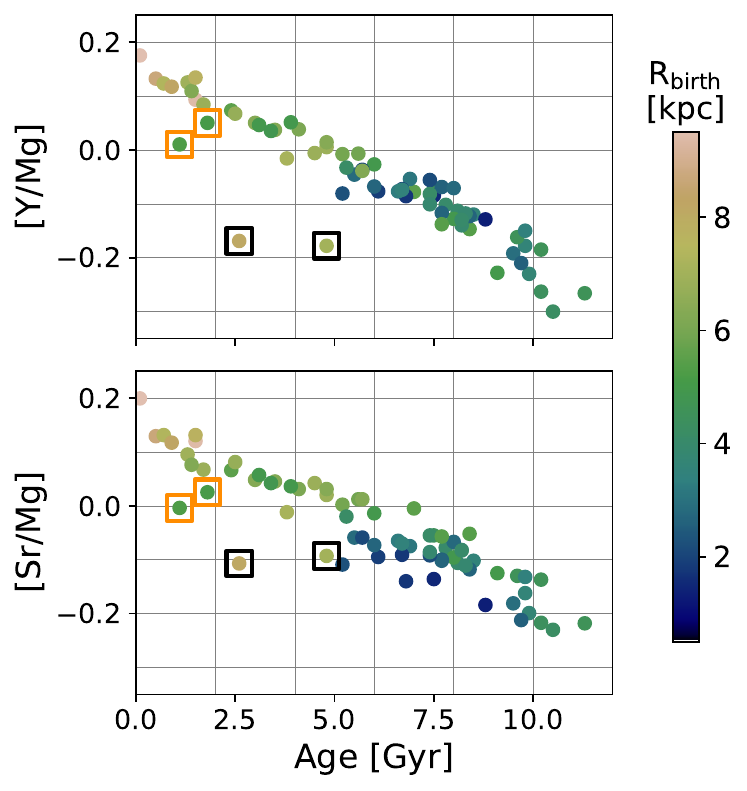}
\caption{\textbf{Top:} The age--[Y/Mg] and \textbf{bottom:} age--[Sr/Mg] relation of 72 nearby solar-like stars presented by \protect\cite{Nissen2020}. Stars are colored by their estimated \rb. The linear [Y/Mg] and [Sr/Mg] relations with age results from stars with different \rb\ across age, with the general trend of younger stars born at larger radii. The main deviants from this line (the components of $\zeta$Retucili (black squares) and the two Na-rich stars (orange squares) are due to their differing birth location compared to other stars at a similar age. This figure illustrates that scatter about the chemical clock relation is a result of a sample containing stars born at different radii at a given age.}
\label{fig:nissen}
\end{figure}

\begin{figure*}
    \centering
     \includegraphics[width=.8\textwidth]{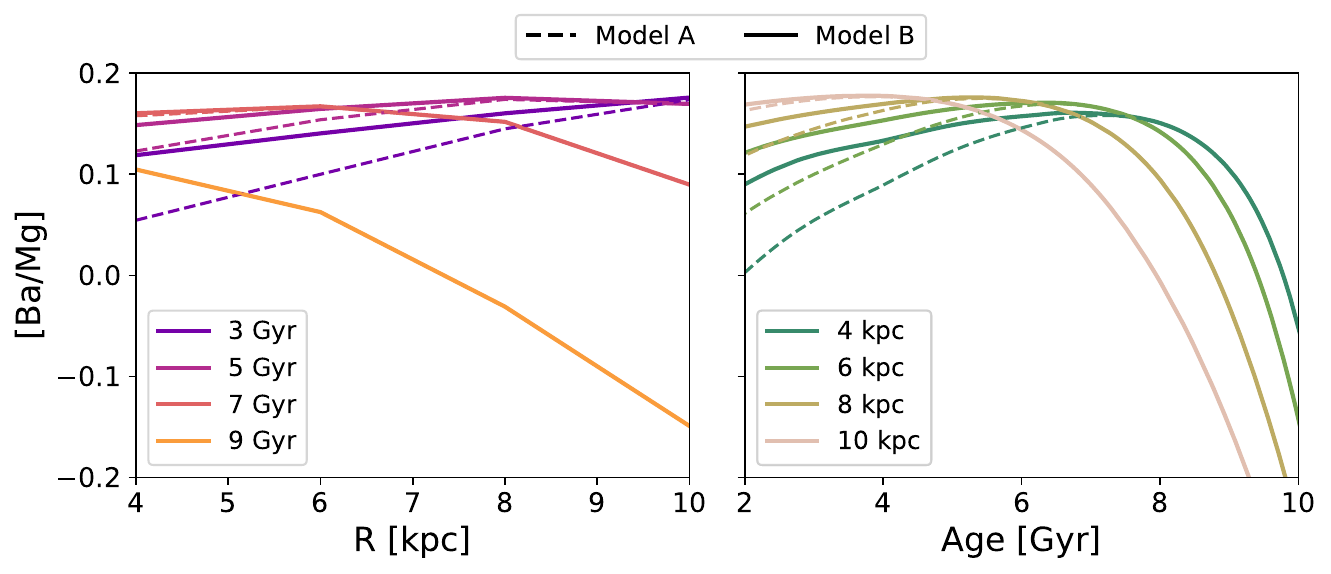}
\caption{The {\bf left:} [Ba/Mg] vs \rb\ and {\bf right:} [Ba/Mg] vs age planes using a single-infall GCE model with inside out growth and simple radial migration as described in Section \ref{sec:model} (Model A, dashed lines). Model A fails to reproduce the observed relationship between age, [Ba/Mg], and \rb. Particularly, the right panel shows that mono-\rb\ populations see a decline in [Ba/Mg] with decreasing age --- unlike the trends shown in Figure \ref{fig:xmg_age}. Model B (solid lines), which allows for more production of Ba at higher metallicities in AGB stars, corrects this issue, and is able to capture the overall trends observed with our \rb. }
\label{fig:model}
\end{figure*}

The tight relation between \smg\ and age found for solar twins in e.g. \cite{Nissen2015, TucciMaia2016, Nissen2020} led to the attempt of expanding this relation to larger samples. While some works have not found a metallicity dependence \citep{Titarenko2019}, others have noted a dependence on \feh\ when examining larger regions of the Galaxy, suggesting that the linear relation between age and chemical clocks is limited by \feh\ \citep{Feltzing2017, DelgadoMena2019}. For example, \cite{Casali2020} found that lower-\feh\ stars had a higher correlation between [Y/$\alpha$] and age, while the higher-\feh\ stars had the weakest slope in the age--[Y/$\alpha$] plane. 

% \im{very good paragraph below!}
Since \rb, \feh, and age are directly linked, it is natural that the differences across metallicity described above are potentially driven by varying \rb. Indeed, for a given age, the higher-\feh\ stars have smaller \rb, which are the birth radii with the weakest age--\smg\ relation. Another way to think of this is that the lower-metallicity stars come from a wider variety of \rb\ (see the right panel of Figure 2 in \citealt{Ratcliffe2023_enrichment}), and therefore run through more mono-\rb\ populations in Figure \ref{fig:xmg_age} faster than the higher-\feh\ stars. The dependency on \feh\ therefore seems to be due to looking at stars born in different places of the Milky Way. 

What does this mean for the samples of solar-like stars that show tight relations between age and \smg? Figure \ref{fig:nissen} shows the age--[Y/Mg] and age--[Sr/Mg] planes for the 72 nearby solar-type stars provided in \cite{Nissen2020}, colored by their derived \rb. It is clear that there is a gradient with \rb\ in these planes for this sample, with younger stars having larger \rb\ and older stars having lower \rb. The minor scatter seen about the correlation is due to varying \rb\ for a given age, which is caused by the variation in \feh\ at that age. In other words, the variation in \smg\ about the chemical clock relation at a given age is due to stars being born at different birth radii. This is especially apparent in [Sr/Mg], which does not show as tight of a correlation with age as [Y/Mg] does.

We find that the four stars that lie off of this tight relation particularly for [Y/Mg] ($\zeta^1$Ret, $\zeta^2$Ret, and the Na-rich stars, marked with black and orange squares respectively) are born at predominantly different locations than the other stars with similar age in this sample.

\subsection{Comparison with Galactic chemical evolution models} \label{sec:model}

To interpret our results, we compare with the age--\smg\ relationship produced using Galactic chemical evolution (GCE) modeling. We use the same nucleosynthesis adopted in the chemical evolution model by \citet{VanderSwaelmen2023} (model D), for what concerns magnesium and the r-process component of barium. The s-process component of barium is the same as assumed in \citet{Rizzuti2019}, with the s-process production of AGB stars taken from the nucleosynthesis prescriptions of \citet{Cristallo2015} and those of massive stars from \citet{Frischknecht2016}. The Galactic disk is described as rings of 2 kpc,  with a single episode of infall following a decreasing exponential law with a fixed timescale of 5 Gyr, not varying with the galactic radius. These are simplified assumptions, compared to a possible double, or even triple, infall (see e.g. \citealt{Spitoni2021, Spitoni2023}) and also to the usual inside-out formation. The remaining parameters of chemical evolution such as initial mass function, star formation law, total surface mass density as a function of galactocentric distance and stellar lifetimes are the same as the disc phase considered in \citet{Cescutti2007}. We use a single infall to naturally obtain a metallicity gradient flattening with time, as found in \cite{Lu2022_Rb} and \cite{Ratcliffe2023_enrichment}. 

Unlike massive stars, AGB stars are long lived and thus have enough time to migrate and pollute neighboring bins. To account for this effect, we implement radial mixing in a simplified manner by assuming that at each time-step a fixed fraction of 10\% of the barium in each ring is transferred to the ring outside. This mimics the enrichment by a percentage of AGB which migrates toward the outskirts of the disk, where they release their s-process enrichment; we define this as Model A. 

% Using a fraction of 20\% provides worst results compared to 10\%; at least with these simplified assumptions, the migration needed is mild.

% In the comparison of the theoretical results of this model and the time evolution of [s/Mg] abundance gradients for our data, we note a discrepancy. The theoretical mono-age predictions show decreasing trends for old ages and only mildly increasing for young ages [showing model A without migration in an appendix?], in contrast with our findings. 

%In Fig.\ref{fig:model}, left panel we show the results of model A. ...
%The results appear closer to the results we obtain based on the data of APOGEE and GALAH. 

%second problem, decreasing in mono rings toward lower ages...
%counteract with a fixed pollution of s-process above certain metallicity. Clearly not expected by theory, maybe the theory has a  shift or longer lifetimes for AGBs? 

%\edits{We make two simple adjustments to the model to account for a metallicity gradient that flattens with time (as found in \citealt{Lu2022_Rb, Ratcliffe2023_enrichment}) and radial migration (by considering the pollution of barium by low mass stars as moving toward the outskirts).} 

The time evolution of the radial gradient of Model A is presented in the left panel of Figure \ref{fig:model} (dashed lines). This model captures the overall trend of the [Ba/Mg] radial gradient across time that we find in Figure \ref{fig:xmg_R}, i.e., the radial gradient begins negative and becomes positive with time. However, the age--[Ba/Mg] relation for Model A (right panel of Figure \ref{fig:model}, dashed lines) shows significant differences from our observational findings. This model predicts that after the initial pollution the [Ba/Mg] abundance stays relatively constant with time for the outer radii, while in the inner disk its abundance significantly decreases (Figure \ref{fig:model}). This is in strong disagreement with our results in Figure \ref{fig:xmg_age}, where we find the [Ba/Mg] abundance in the inner disk stays fairly constant with age and the outer disk sees a steep increase with time.

% The radial gradients for each mono-age population are shown to be primarily linear, in slight disagreement with the non-linear trends shown in Fig. \ref{fig:xmg_R}, left panel. They do however capture the originally negative gradient, which transitions to a positive gradient that steepens with time. 

%The age--[Ba/Mg] relation for Model A though shows significant differences from our observational findings, where here each mono-\rb\ population sees a strong decrease in [Ba/Mg] with time, particularly for the inner radii. This decrease also affects the overall age--[Ba/Mg] trend, where the negative correlation between [Ba/Mg] and age becomes quadratic. 

%The left and right panels of Figure \ref{fig:model} show the expectations of our GCE model (denoted ``Model A"; dashed lines) in the \rb--[Ba/Mg] and age--[Ba/Mg] planes respectively. 

%\edits{One explanation for the decrease in [Ba/Mg] with time produced in Model A is that the amount of Ba produced in higher metallicity AGB stars is too little. To correct for this, we run Model A with a modification to the yields; the Ba yields for $Z > 0.01$ are replaced with the yields of $Z=0.01$.} The expectations of this modified model (Model B) are shown as the solid lines in Figure \ref{fig:model}. We see that the discrepancies listed above are primarily resolved; that is, the [Ba/Mg] abundance at each radii stays relatively constant with time after its initial increase and the \rb--[Ba/Mg] trends are non-linear for each mono-age population. 

A possible explanation for the decrease in [Ba/Mg] with time produced in Model A is that the amount of Ba produced in the models of high metallicity AGB stars is too low. To analyze this solution, we run Model A with a modification to the yields; the Ba yields for $Z > 0.01$ are replaced with the yields of $Z=0.01$. In our set of nucleosynthesis, the yields of barium at $Z=0.01$, are about a factor of 2 larger than the yields at $Z=0.02$, depending on the mass of the AGB considered.
The expectations of this modified model (Model B) are shown as the solid lines in Figure \ref{fig:model}. We see that the discrepancies listed above are primarily resolved; that is, the [Ba/Mg] abundance at each radius stays relatively constant with time after its initial increase. The results of Model B also appear to agree with what concerns the [Ba/Mg] vs \rb\ plot (solid lines in left panel of Figure \ref{fig:model}), although in this plane the difference imposed by the variation in the yields are less significant. We underline that with our data it is not trivial to disentangle possible enrichment of s-process elements happening in binary systems.  In this scenario, if the primary ended its life at AGB, the secondary star could be enhanced in s-process elements; at the same time, the mass transfer can bias the determination of the age toward younger ages. Further investigations are needed to  understand the impact of this population that at this stage we consider negligible.

%We acknowledge that neither model perfectly reproduces the data due to the simplifications and (many) unknown parameters in modeling the Milky Way's evolution. While the choice of how to replace the higher metallicity yields is more or less arbitrary, the goal of this section is to show [the limitations of a standard GCE model, and] how we are able to capture the main trends from Figures \ref{fig:xmg_R} and \ref{fig:xmg_age} with a simple modification. 

The results that we found for [Ba/Mg] can be explained by a disk with a decreasing exponential surface density and with a fixed infall timescale of 5 Gyr for all the radii, where the inner part enriches faster due to higher star formation efficiency, in an inside-out fashion. This combined with a strong metallicity dependence in the yields of s-process, with a mild modification of the most metal-rich tail, and with a migration of around 10\% of the AGB products can explain the trends found in the data. The results of this section are robust to different values of this migration percentage, with a smaller/larger fraction causing smaller/larger variations in \smg\ across \rb\ for younger ages. Choosing the optimal fraction is nontrivial and outside the scope of this work. However, at least with these simplified assumptions, the migration needed is mild. 

We acknowledge that the model does not perfectly reproduce the data as there are many parameters in modeling the Milky Way's evolution. Moreover, the choice of how to replace the higher metallicity yields is arbitrary. Still, the goal of this section is to show that this model is able to capture the main trends from Figures \ref{fig:xmg_R} and \ref{fig:xmg_age} and it provides us an interpretation of our results. This section also shows the power of \rb, and how it can be used to constrain GCE models.

\section{Conclusions} \label{sec:conclusion}
% past tense 

This work investigated the age--\smg\ relation across \rb\ using \numApp\ APOGEE DR17 red giant disk stars and \numGal\ GALAH DR3 MSTO+SGB disk stars. Previous works found variations in chemical clocks with current radius, however due to stars radially migrating away from their birth sites, these fundamental trends can be difficult to interpret. In this work, we explore the relative evolution of s-process and $\alpha$-elements across time with birth radii. Our key findings are: 

\begin{itemize}

    \item The \smg\ radial birth gradient for all three elements examined started initially weakly negative, flattened around the time of high- to low-$\alpha$ transition, and then became increasingly positive (or concave downwards) toward redshift zero (Figure \ref{fig:xmg_R}).
    
    \item The age--\smg\ relation varies across \rb, with each radius seeing an initial increase in \smg\ before varying little with time (Figure \ref{fig:xmg_age}). This suggests that the scatter about the \smg--age line is predominantly due to samples containing a variety of \rb\ for a given age. 
    
    \item When the dispersion in \rb\ is small for a given age (such as in Milky Way solar-like stars), a tight correlation between \smg\ and age is found (Figure \ref{fig:nissen}). 

    % \item Using a GCE model, we find that in addition to modeling radial migration, more Ba needs to be produced in higher metallicity AGB stars to reproduce the trends found using our \rb\ estimates (Figure \ref{fig:model}). 
    
    \item Using a GCE model with standard nucleosynthetic yields, we find that the [Ba/Mg] abundance decreases with time for each mono-\rb\ population in disagreement with the data. A simple correction for allowing more Ba to be produced at higher metallicities, as well as a simple model of radial migration, address this issue and is able to reproduce the trends found using our \rb\ estimates (Figure \ref{fig:model}).

\end{itemize}

Our results explicitly show that there is no universal \smg--age relation across the Galactic disk, in agreement with previous works \citep[e.g.][]{ViscasillasVazquez2022}. Here, however, we revealed how this correlation depends on \rb\ (and thus \feh), which demonstrates the inherent relationship between \smg\ abundance and age in the Milky Way that is not masked by radial migration. In particular, we illustrate how radial migration affects the \smg\ radial gradient with time, which in turn creates a variation in the \smg\ abundance across birth radii and causes a weaker correlation between age and \smg\ for younger ages.

The time evolution of the age--\smg\ relation that we uncovered with knowledge of birth radius was used to constrain the metallicity dependence of the AGB yields, and the radial migration strength of a simple GCE model --- which is typically constrained by only present-day observations with age and radius --- and highlights the utility of \rb\ in understanding the Milky Way's chemical evolution.

\section*{Acknowledgements}
 
B.R. and I.M. acknowledge support by the Deutsche Forschungsgemeinschaft under the grant MI 2009/2-1. This work was also partially supported by the European Union (ChETEC-INFRA, project no. 101008324)

%%%%%%%%%%%%%%%%%%%%%%%%%%%%%%%%%%%%%%%%%%%%%%%%%%
\section*{Data Availability}

The datasets used and analysed for this study are
derived from data released by APOGEE DR17, GALAH DR3, \cite{Queiroz2023_SH} (corrected version: \href{https://data.aip.de/projects/aqueiroz2023.html}{https://data.aip.de/projects/aqueiroz2023.html}) , \cite{Anders2023_ages}, and \cite{Nissen2020} \citep{Nissen2020_data}. The rest of the relevant datasets are available from the corresponding author on reasonable request.

% \appendix
% \label{sec:append}

% \section{Extra Plots} 

\bibliographystyle{mnras}
\bibliography{Ratcliffe}{}

\bsp	% typesetting comment
\label{lastpage}
\end{document}